\documentclass[twocolumn,showpacs,prb,longbibliography]{revtex4-2}
\usepackage{graphicx}
\usepackage{color}
\usepackage{dcolumn}
\usepackage{bm}
\usepackage{mathtools}
\usepackage{float}
\usepackage{hyperref}
\usepackage{footnote}
\usepackage[normalem]{ulem} 
\usepackage {ulem} 

\begin{document}
\title{Excess soft photon production in $Z^o$ decay into hadronic jets}
\author{Giulia Pancheri}
\affiliation{Laboratori Nazionali di Frascati, INFN, Frascati, Italy}
\author{Yogendra Srivastava}
\affiliation{Emeritus Professor of Physics, Northeastern University, Boston MA USA\\ 
and \\
Associate, Centro Ricerca Enrico Fermi, Roma, Italy\\
and\\
INFN Sezione di Perugia, Perugia, Italy
}

\begin{abstract}
In view of a long standing discrepancy between the single soft photon spectrum computed using Haissinski-Low formulae when compared with the 
experimental results, we reconsider a simpler problem: the bremsstrahlung spectrum associated with hadronic jets detected through decays of a 
resonance (produced say in electron-positron collisions). Our  proposed explanation for the excess number of anomalous  soft photons observed by 
the DELPHI collaboration in 2-jet decays of the $Z^o$ is related to the existence of a finite temperature for confined quarks. We calculate the excess 
number of very soft photons in good agreement with DELPHI’s measurement with our  temperature range of [$k_BT= (130-152) MeV$].  
\end{abstract}

\maketitle

\section{Introduction \label{intro}}

It has been known for a few decades that there is an excess in the number of photons produced in association with hadrons (but not leptons) in high 
energy processes; the excess being defined with respect to the bremsstrahlung spectrum computed via standard Haissinski-Low formulae in QED \cite{Low:1958sn,Haissinski:1987jb,Fadin:2024tar}.
 For example, photons produced in association with lepton pairs ($l^+l^-$, $\bar{l}\nu$ or 
$l\bar{\nu}$) produced in decays of a $J/\Psi$, $W^{\pm}$ or a $Z^o$ resonance show no excess 
\cite{Fleischer:1984vt,FermilabE760:1996jsx,Spiridonov:2004mp,DELPHI:2007nmh} 
whereas there is an {\it excess} in the number of soft photons produced in association with hadronic decays of the 
same resonances, see for example  \cite{DELPHI:2005yew}. \\
\\

The subject of this photon {\it anomaly} has been  thoroughly reviewed in a recent Physics Report \cite{Bailhache:2024mck},  where an extensive set 
of experimental references and theoretical suggestions can be found. Explanations  offered towards this excess, include  it  as being due to a novel coherent hadronic state 
 \cite{Barshay:1994nh} and  relating  soft photon anomalies  to  Quark Gluon Plasma effects on QED bremsstrahlung  \cite{Shuryak:1978ij,Lichard:1990ye,PhysRevC.89.044910,Volkl:2026fuw}.\\ 
\\
This excess was observed in  hadronic collisions through a large range of initial state energies and different experimental final states.  As the puzzle persists,  the ALICE collaboration has included the study of the anomalous soft photon effects in a letter of intent \cite{KarlssonGumprecht:2026fuo,ALICE:2022wwr}.\\
\\
In this note, we shall argue that the observation of the effect in leptonic reactions, such as the one reported by the DELPHI collaboration at LEP, offers the possibility to quantitatively connect the effect to
 the temperature of the confined state of quarks created by $Z^o$-decays, before the evolution into hadronic jets and, in  the next section, we shall offer our own rather modest proposal for this excess based on two simple observations. First is based on the Bose-Einstein 
notion of photon statistics: if in a quantum state there are (already) present $N_\Omega$ photons (of a given frequency $\Omega$), then the probability 
to add an extra photon (of the same frequency $\Omega$) is increased by a factor ($N_\Omega+1$). That is, in the properly normalized number representation 
($|n>$) of photons, the action of a single photon creation operator ($a^\dagger$) has the following well-known form:
\begin{eqnarray}
\label{b1}
a^\dagger(\Omega) |N_\Omega> = \sqrt{N_\Omega +1}\ |N_\Omega+1> ;(i)\nonumber\\
{\rm Thus, the\ mean\ number\ of\ photons\ after\ addition}   \nonumber\\
{\rm of\ another\ photon\ of\ the\ same\ frequency}:\nonumber\\
\mathcal{N}_\gamma (\Omega)= <N_\Omega| a_\Omega a^\dagger_\Omega |N_\Omega> = (N_\Omega +1);(ii)\nonumber\\
\end{eqnarray}
an utterly unsurprising result.\\
\\
Hence,  it follows that if there is an excess in the mean number of photons produced in association with hadrons compared to the mean number of photons produced in 
association with leptons (both final states produced from the decay of {\it the same resonance}), it necessarily implies that there must have been {\it extra} 
(number $N_\Omega$ of) photons already present -from some dynamical mechanism yet to be specified- when the resonance under study decayed into 
hadrons but such {\it extra} photons were absent in its leptonic decays.\\
\\
However, the mechanism producing these extra photons when the decays are in the hadronic mode but never in the leptonic mode must also be elastic 
enough to cover a production mechanism that is over a large enough band of photon energies as the hadronic anomaly persists over a sizable mass range 
(from say bottomonia to $Z^o$) of hadronic resonances. Otherwise said, the source of extra photons must cover a very large (perhaps extending up to infinite) 
energy range. This leads us naturally to our second notion: that of the Planck radiation formula -{\it primogenitore} of quantum mechanics- that is based on 
an infinite photon frequency spectrum ($0\leq \Omega \leq \infty$) at any given temperature ($T$): the spectrum at a given temperature ($T$) cannot arbitrarily 
be truncated at an arbitrary small but finite ($\Omega_{min}$) or, any however large but finite frequency ($\Omega_{max}$) without introducing inconsistencies 
or discontinuities in the source that is by hypothesis being maintained at a given homogeneous temperature ($T$) \footnote{It is precisely the deviation from 
the pristine Planck radiation formula in the large scale survey of the early Universe, that led astrophysicists to interpret local variations in temperatures called 
{\it hot spots} implying cosmic mass inhomogeneities. For the problem under consideration on the other hand, it should be a reasonable working hypothesis 
to assume  a homogeneous ambient temperature at least during the hadronic fragmentation of a confined (light mass quark).}.\\
\\
In short, Planck distribution at a specified temperature requires a distribution over all frequencies. Hence, our quest is limited to a search of a fixed temperature 
($T>0$) source for the hadronic mode of our resonances but not for its leptonic modes. Just such a scheme realizing this requirement would be considered and 
amplified upon in the next Sec.(\ref{excess}). A central point of our proposal is to identify  the full width at half maximum of the resonant state decaying into hadronic jets 
as the intrinsic parameter defining the volume of the confined region where the observed soft photons can be radiated \cite{Bjorken:1979hv}.

\section{A simple thermodynamic model for excess photons in hadronic decays \label{excess}}
Let us consider the long standing puzzle discussed in the previous Sec.(\ref{intro}) about an {\it excess} in the number of photons produced in association 
with hadrons through the concrete example of the Delphi experiments performed at LEP\cite{DELPHI:2007nmh,DELPHI:2005yew}. In a set of two experiments 
performed by the Delphi Collaboration at the LEP $e^+e^-$ Collider, soft photons were observed both in the $\mu^+\mu^-$ lepton pair as well as in the hadronic 
jet channels arising from the decays of the neutral boson $Z^o$. When standard soft-photon expressions were used, agreement between data and theory was good 
for the lepton channel but there was a clear excess in the number of soft photons in the hadronic jets. To keep our discussion as simple as possible, let us consider 
the final state of 2 jets (that is a $q\bar{q}$ jet)\\
\\
At first glance, to lowest order in $\alpha$ in QED, apart from a numerical change in the value of the electric charges in ($l^+l^-$) and ($q\bar{q}$) channels, 
one expects hardly any change in the manner of soft photon production between the two cases\cite{Low:1958sn,Etim:1967mxo,Haissinski:1987pk}. However, 
such a view is obviously too simplistic because while leptons can be assumed to be produced essentially in the vacuum, clearly the quarks are not as they are 
{\it confined}. In fact, in a very early paper \cite{Grillo:1979wu}, it was noted that under very general conditions, every confined system must develop an intrinsic 
temperature. Using the MIT bag model for confinement of quarks, an estimate of $k_BT_{conf} \approx\ 130$\ MeV was made in reference \cite{Hosoya:1978rj}. 
In the following, we shall estimate the total number of {\it thermal} photons produced from a source such as in the two-jet ($q\bar{q}$) events of the Delphi experiment 
\cite{DELPHI:2005yew} provided by $Z^o$ decays assumed to be at an intrinsic temperature $T$.  \\
\\       
The {\it energy} density (energy per unit volume) of photons at a temperature ($T$) in a frequency interval ($d\nu$) is given by the Planck distribution 
(see, for example \cite{Weinberg:1972kfs}) 
\begin{eqnarray}
\label{P1}
\rho_\gamma(\nu;T) d\nu= (\frac{8\pi}{c^3}) \frac{(h\nu)\nu^2 d\nu}{[e^{(\frac{h\nu}{k_BT})}-1]}.
\end{eqnarray}
In terms of $h=2\pi \hbar$ and the angular frequency $\omega=2\pi \nu$, we have
\begin{eqnarray}
\label{P2}
\rho_\gamma(\omega;T) d\omega= (\frac{1}{\pi^2 c^3}) \frac{(\hbar \omega)\omega^2 d\omega}{[e^{(\frac{\hbar\omega}{k_BT})}-1]}.
\end{eqnarray}
Thus, the {\it number} density of photons at temperature ($T$) per unit angular frequency ($d\omega$) per unit volume is given by
\begin{eqnarray}
\label{P3}
n_\gamma(\omega;T) d\omega= (\frac{1}{\pi^2 c^3}) \frac{\omega^2 d\omega}{[e^{(\frac{\hbar\omega}{k_BT})}-1]}.
\end{eqnarray}
Hence, the total number of photons per unit volume [$n_\gamma(T)$] is given after an integration over all photon energies ($E=\hbar \omega$) by
\begin{eqnarray}
\label{P4}
n_\gamma(T) = \Big{(}\frac{1}{\pi^2 (\hbar c)^3}\Big{)}\int_o^\infty \frac{E^2 dE}{[e^{(\frac{E}{k_BT})}-1]};(i)\nonumber\\
=(\frac{1}{\pi^2}) \Big{(}\frac{k_B T}{\hbar c}\Big{)}^3 \int_o^\infty \frac{x^2 dx}{[e^x -1]};(ii)\nonumber\\
= (\frac{1}{\pi^2}) \Big{(}\frac{k_B T}{\hbar c}\Big{)}^3 \Gamma(3) \zeta(3),(iii)
\end{eqnarray}
where the Gamma function $\Gamma(3)=2$ and the Riemann zeta-function $\zeta(3)\approx\ 1.2$.\\
\\
Thus, the total number of photons [$N_\gamma(T,V)$] in a volume ($V$) (after integrating over all photon energies and over all angles) 
\begin{eqnarray}
\label{P5}
N_\gamma(T, V) = (\frac{2 V}{\pi^2}) \Big{(}\frac{k_B T}{\hbar c}\Big{)}^3 \zeta(3).
\end{eqnarray}
To proceed further, the fiducial volume occupied by the photons observed in $Z^o$-decay into jets needs to be defined.    
For matter and  radiation {\it in equilibrium}, we can estimate the fiducial volume $V$ in $Z^o$ decays into jets as follows. The full width at half-maximum 
$\Gamma$ of $Z^o$ defines a time $\tau=(\hbar /\Gamma)$, and a length $\Lambda=(hc/\Gamma)=(2\pi \hbar c)/\Gamma$ and thereby a volume 
$V_o=(4\pi/3)\Lambda^3$ (for an isotropic decay). Putting it all together, we have
\begin{eqnarray}
\label{P6}
V_o= (\frac{32\pi^4}{3}) (\frac{\hbar c}{\Gamma})^3;(i)\nonumber\\
N_\gamma(T, V_o) = (\frac{64 \pi^2}{3}) \Big{(}\frac{k_B T}{\Gamma}\Big{)}^3 \zeta(3);(i)\nonumber\\
\approx\ (0.4 \pi^2)\Big{(}\frac{4 k_B T}{2.5\ GeV}\Big{)}^3;(ii)
\end{eqnarray}
We advocate the choice of  $\Gamma$, the full width at half maximum, to define the volume,  as  consistent with it being the intrinsic parameter 
which regulates radiative corrections to narrow resonance production in electron positron collisions \cite{Pancheri:1969yx,Greco:1975ke}.\\
\\
The estimate of the intrinsic temperature was made as follows in  \cite{Grillo:1979wu,Hosoya:1978rj}. For a linearly rising (confining) potential, 
$V(r)=\sigma  r$, where $\sigma$ is the string tension, we have a constant acceleration $a= \sigma/m_q$, where $m_q$ is the (constituent) mass of a 
light quark confined in a $\rho$-meson or a nucleon. In this model, the string tension can be related to the Regge trajectory slope $\alpha^{'}$ 
(in units of $GeV^{-2}$),  so that 
\begin{eqnarray}
\label{Q1}
\sigma = \frac{1}{4 \hbar c \alpha^{'}};(i);\nonumber\\
(k_B T) = \frac{\hbar a}{2\pi c} = \frac{1}{8\pi \alpha^{'} m_q c^2};(ii);\nonumber\\
\end{eqnarray}
A few words are in order as to our choice of the linear confining potential, $V(r)= \sigma  r$. Only for this potential, we have (i) a constant acceleration 
$a=\sigma/m_q$, and thus a rigorous -thermodynamic connection between temperature ($T$) and the acceleration ($a$): 
$k_B T= (\hbar a)/(2\pi c)$ \cite{Grillo:1979wu,Hosoya:1978rj}; (ii) a linearly rising Regge trajectory $J=J_o + \hbar \alpha^{'} (Mc^2)^2$ and 
(iii) an equipartition rule (for light mass quarks) between rotational and confinement energies leading to the relation $\sigma= 1/(4 \hbar c \alpha^{'})$, \cite{PhysRevD.63.077502}.

It is interesting to note that (the right hand side of) Eq.(\ref{Q1}(ii)) is purely {\it classical} as there is no $\hbar$ in it. For a numerical estimate, in \cite{Hosoya:1978rj}, 
$m_q= 300$ MeV and $\alpha^{'}= 1\ GeV^{-2}$ were used, leading to an intrinsic temperature of $130$ MeV. Whereas if we employ a more 
realistic estimate of $\alpha^{'}= (0.87-0.88)\ GeV^{-2}$, we obtain $k_B T= (150-152)$ MeV, closer to  Hagedorn temperature of ($158\pm 3$) MeV 
\cite{Hagedorn:1965st,Broniowski:2000bj}. \\
\\
Thus, for a linearly rising potential, we reach the following estimate
\begin{eqnarray}
\label{Q2}
(k_B T) = (130-152) MeV
\end{eqnarray} 
with the upper limit close to  recent  estimates from lattice calculations \cite{Giordano:2026bqj}.\\
 Proceeding further to obtain concrete numerical results, let us assume the (lowest) nominal value
$k_B T=0.13$\ GeV  \cite{Grillo:1979wu,Hosoya:1978rj}. For it,
we find for the total number of $\gamma$'s,
\begin{eqnarray}
\label{P7}
N^{linear}_\gamma(T, V_o)\approx\ (35.5 \times 10^{-3})\times \Big{(}\frac{ k_B T}{0.13\ GeV}\Big{)}^3 
\end{eqnarray}
produced over all (soft and hard) photon energies and over all angles. To compare with data, we have to apply appropriate cuts in angle and energy of the observed
photons.\\
\\
Before analyzing the DELPHI data on $Z^o$ two-jet events with kinematic cuts, let us point to a general ({\it scaling}) result. From Eq.(\ref{P6}), we see that for any 
resonance decaying via 2-jets, the total number of photons scales as ($(k_B T/\Gamma)^3$) where $T$ is the temperature and   $\Gamma$ is the partial width of the 
resonance into the chosen quark channel. Thus, the total number
 of {\it anomalous} photons in Higgs decay via the ($H\to b\bar{b}$) channel should provide important information both about the temperature and the Higgs partial decay 
 width into this channel. The theoretical analysis of Higgs decay \cite{Fox:2026tkh} is expected to be explored through the   HL-LHC programme \cite{ATLAS:2025eii}. 
 The present  discussion of the $Z^0$ case shows that their explorations at future accelerators, such as a possible  muon collider  \cite{Andreetto:2026jxm,Palmer:2026nal}, 
 could be of special interest, as are   precise radiative corrections to Higgs production and decay \cite{ Greco:2015yra,Franceschini:2021aqd}. \\
\\    
Let us now return to the 2-jet decays of $Z^o$ and estimate the kinematic cuts to compare our theoretical results to experiment.\\
\\
{\bf I: cuts in angle:}\\ 
Delphi photons are included only for angles [$0.01(\rm radians)\leq \vartheta_\gamma\leq 0.4(radians)$] where $\vartheta_\gamma$ is the polar angle between
the observed photon and the jet axis. As the thermal radiation is isotropic, for each jet, we should have, for small angles ($\vartheta_{1,2}$), the following angular cut
reduction factor
\begin{eqnarray}
\label{P8}
(\Delta \vartheta)_{cut} \equiv\ (\frac{1}{2})\int_{cos\vartheta_2}^{cos\vartheta_1} dcos(\vartheta_\gamma);(i);\nonumber\\
\approx\  (\frac{1}{2})\int_{0.01}^{0.4} \vartheta_\gamma (d\vartheta_\gamma)\approx\ 0.04;(ii);\nonumber\\
\end{eqnarray}
Applying the above angular cut, we find for the number of $\gamma$'s  
to be:
\begin{eqnarray}
\label{P9}
N_\gamma({\rm all}\ E_\gamma)/jet \approx\ (1.42 \times 10^{-3})\times \Big{(}\frac{ k_B T}{0.13\ GeV}\Big{)}^3.
\end{eqnarray}
\\
{\bf II: cuts in energy:} Delphi photons are required to be in the energy range [$0.2\ GeV \leq E_\gamma \leq 1\ GeV$]. This implies that the Riemann zeta function $\zeta(3)$
should be replaced in Eq.(\ref{P6}) by
\begin{eqnarray}
\label{P10}
\zeta(3) \to\ P(x_1;x_2)= (\frac{1}{2}) \int_{x_1}^{x_2}(\frac{x^2 dx}{[e^x -1]}), 
\end{eqnarray}
where $x_1=(0.2\ GeV)/(k_B T)\approx\ 1.54$ (for the nominal value of $k_B T= 0.13$\ GeV) and $x_2=(1\ GeV)/(k_B T)\approx\ 7.7$ (for the nominal value of $k_B T= 0.13$\ GeV).
As $e^{7.7} \approx\ 4.6\times 10^4$, we may as well set $x_2 \to \infty$ without incurring much error. Pending a more accurate calculation (not required at the present state of analysis of the data), 
our present estimate (based on the numerical tables in \cite{Stegun:1964}) is
\begin{eqnarray}
\label{P11}
P(x_1=1.54;x_2=\infty) \approx\ 0.85
\end{eqnarray}
Including the above energy cut reduction, our final estimate for the number of (thermal) photons satisfying the Delphi energy and angle cuts, per jet, to be approximately $0.85/1.2=0.70$
of the value found in Eq.(\ref{P9}), for $k_B T=0.13$\ GeV:
\begin{eqnarray}
\label{P12}
(\frac{N_\gamma}{jet})|_{Delphi-cuts} \approx\ (1.01 \times 10^{-3}) \times \Big{(}\frac{ k_B T}{0.13\ GeV}\Big{)}^3.   
\end{eqnarray}
Thus, for the  thermodynamical model discussed here with the confinement temperature lying between $k_B T=(130-152)$ MeV, with the Delphi cuts as applied above,
we expect the number of photons per jet to lie in the interval
\begin{eqnarray}
\label{P13}
(\frac{N_\gamma}{jet})|_{Delphi-cuts} \approx\ (1.01-1.61) \times 10^{-3} .   
\end{eqnarray}

\section{Conclusions and future prospects \label{con}}
It is difficult for us to compare the result above in Eq.(\ref{P12},\ref{P13}) with the Delphi data, in light of the comment in their Abstract \cite{DELPHI:2005yew} that:\\
\\
{\it This excess (uncorrected for the photon detection efficiency) is $(1.17 \pm 0.06 \pm 0.27) \times 10^{-3}$$\gamma$'s/jet in the specified kinematic region, while the
expected level of the inner hadronic bremsstrahlung (which is not included in the Monte Carlo) is $(0.340 \pm 0.001 \pm 0.038) \times 10^{-3}$$\gamma$'s/jet.}\\
\\
It is  unclear how radiative corrections to $Z^o\rightarrow 
 \mu^+\mu^-$  were applied, namely whether the MonteCarlo included, for instance,  results such as from \cite{Greco:1980mh}. We can thus interpret the above comment in two ways:\\
\\
(i) That the standard theory gives ($0.34\times 10^{-3}$) photons and Delphi finds ($1.17\times 10^{-3}$) photons, so that one has to find an extra number ($0.83\times 10^{-3}$) of photons by some
mechanism. This can be done by lowering $k_B T$ to  $k_B T =0.122$\ GeV, a small adjustment to our previous model value for the intrinsic temperature range ($0.13-0.152$\ GeV).\\
\\
(ii) That Delphi finds ($1.17\times 10^{-3}$) photons above and beyond the standard value ($0.34\times 10^{-3}$) photons. In this case, we have to raise the limit in Eq.~(\ref{P3})
for the temperature to
$k_B T=0.137$\ GeV, that is well within our theoretical model range $k_B T= (0.130-0.152)$\ GeV \\
\\
By the way, the above statements are only true if the standard mechanism does not interfere with the thermal mechanism, and if radiative corrections to the leptonic decay mode 
have included resonance effects \cite{Greco:1980mh}.\\
\\
In any case, based on our analysis, it is fair to say that measurements of the number of photons per jet  may be a rather precise way of determining the intrinsic temperature 
as the number of thermally produced photons is proportional to the cube of the intrinsic temperature and that precise radiative corrections to low energy processes may give important clues to this quest \cite{WorkingGrouponRadiativeCorrections:2010bjp,Dave:2026cqi}.   Possible applications of the above semi-classical discussion could also shed 
light on the role of the intrinsic temperature in  radiative meson decays,
where experimental data show discrepancies \cite{DiPalma:2025iud}.
\bibliographystyle{apsrev4-1}
\bibliography{Excess-photons-QCDtempe-lia}
\end{document}